\documentclass[fleqn,twoside]{article}
\usepackage{espcrc2,amssymb}

\usepackage{graphicx}
\usepackage[figuresright]{rotating}

\newcommand{\AmS}{{\protect\the\textfont2
  A\kern-.1667em\lower.5ex\hbox{M}\kern-.125emS}}

\title{Star Formation, Massive Stars, and Super Star Clusters in Nearby 
Galaxies with the SKA}

\author{Kelsey. E. Johnson\address[MCSD]{Dept. of Astronomy, 
    University of Virginia\\ 
        P.O. Box 3818, Charlottesville, VA, 22903, USA}
        \thanks{K.E.J. gratefully acknowledges support for this work 
	  provided by the NSF through an Astronomy and Astrophysics 
	  Postdoctoral Fellowship.}}
       
\begin{document}

\begin{abstract}
The Square Kilometer Array (SKA) will enable studies of star formation
in nearby galaxies with a level of detail never before possible
outside of the Milky Way.  Because the earliest stages of stellar
evolution are often inaccessible at optical and near-infrared
wavelengths, high spatial resolution radio observations are necessary
to explore extragalactic star formation.  The SKA will have the
sensitivity to detect individual ultracompact HII regions out to the
distance of nearly 50~Mpc, allowing us to study their spatial
distributions, morphologies, and populations statistics in a wide
range of environments.  Radio observations of Wolf-Rayet stars outside
of the Milky Way will also be possible for the first time, greatly
expanding the range of conditions in which their mass loss rates can
be determined from free-free emission.  On a vastly larger scale,
natal of super star clusters will be accessible to the SKA out to
redshifts of nearly $z \sim 0.1$.  The unprecedented sensitivity of
radio observations with the SKA will also place tight constraints on
the star formation rates as low as $1 M_\odot$~yr$^{-1}$ in galaxies
out to a redshift of $z \sim 1$ by directly measuring the {\it
thermal} radio flux density without assumptions about a galaxy's
magnetic field strength, cosmic ray production rate, or extinction.

\vspace{1pc}
\end{abstract}

\maketitle

\section{INTRODUCTION}

\subsection{The Importance of Studying Extragalactic Star Formation}

Star formation is a critical process -- arguably one of the most
fundamental physical processes (next to gravitational collapse) in
determining the appearance and properties of the visible universe.
The formation of massive stars is particularly important because
they have a major effect on the energetics of galaxies: massive stars
are responsible for the ionization of the interstellar medium, their
stellar winds and supernovae are main sources of mechanical energy,
their ultraviolet radiation powers thermal-infrared (IR) luminosities
through the heating of dust, they are a main driver of chemical
evolution via supernova explosions at the end of their lives, and they
may be the sources of gamma ray bursts.

Nevertheless, despite the significant role of massive stars throughout
the universe, their birth is not well understood and we are only
beginning to piece together a scenario for the youngest stages of
massive star evolution.  There has been some progress understanding the
early stages of massive star formation in the Milky Way, but the
current knowledge about the early stages of massive star evolution in
other environments is mediocre at best.

The reasons for this dearth of information about extremely young
massive stars in other galaxies are predominantly threefold: (1) the
earliest stages of stellar evolution are deeply enshrouded and
inaccessible at optical and near-infrared wavelengths; (2) compared to
other types of radio sources (e.g. AGN, SNe, SNR), individual stars
are relatively faint and require exceptionally sensitive observations;
and (3) high spatial resolutions are necessary to disentangle the
individual massive stars from their surrounding environment and
background contamination.

The study of star formation in environments outside of our own Milky
Way is critical for advancing our understanding of stellar and
galactic evolution.  The Square Kilometer Array (SKA) will
clearly enable investigations of star formation in the Milky Way with
an unprecedented level of detail.  However, without similar
investigations in other galaxies, it is difficult to place Galactic
work in a larger context.  The SKA will revolutionize the field
of star formation by enabling detailed studies that have never been
possible outside of the Milky Way.

The ability to observe star formation taking place in a range of
conditions will enable us to better interpolate and extrapolate
details of the star formation process throughout the universe.  For
example, we do not know how the properties of star formation depend on
various environmental parameters, including metallicity, pressure,
turbulence, stellar densities, triggering scenarios (including bars,
bubbles, and galactic interactions), or how star formation might
differ in nuclear regions or in ``burst'' and quiescent modes.  
Investigations of all of these issues will be possible with the sensitivity
and resolution of the SKA.

\section{THE NATURE OF STAR FORMATION ON A SMALL SCALE}

\begin{figure*}[htb]
\centerline{
\includegraphics*[scale=0.5]{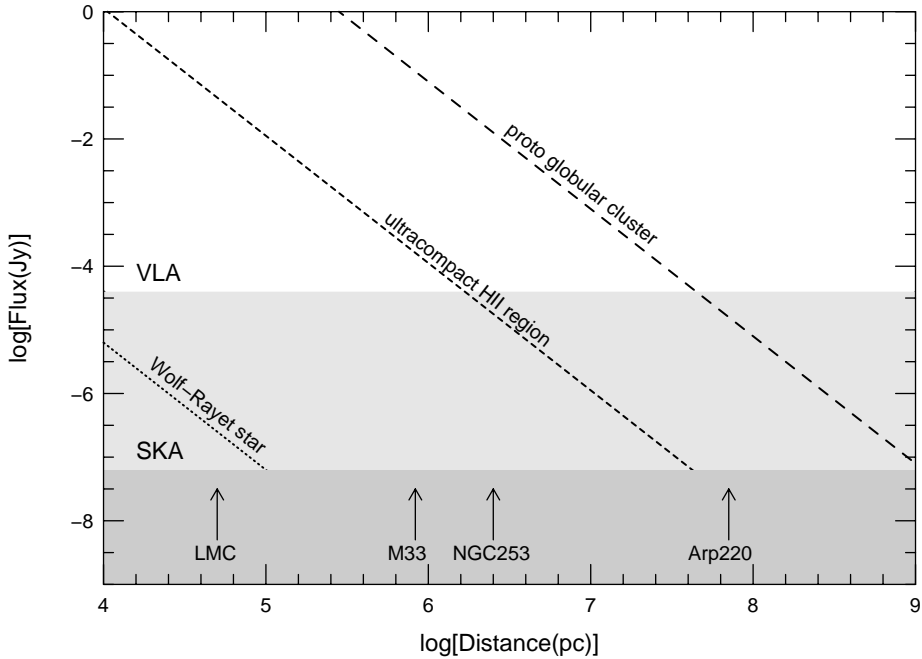}
}
\caption{A schematic comparing the $5\sigma$ sensitivities for the
Very Large Array and Square Kilometer Array for an 8 hour
integration at 6~cm ($\sim 40\mu$Jy and $0.07\mu$Jy, respectively).  The
flux densities of ``typical'' WR-stars, UCH{\sc ii}s, and natal globular
clusters are indicated.
 \label{VLA_SKA}}
\end{figure*}

Because of their extremely short lifetimes, massive stars trace the
most recent episodes of star formation in a galaxy, and therefore
allow a detailed probe of recent star formation history.  In
particular, ultracompact H{\sc ii} regions indicate the {\it current}
($\lesssim 1$~Myr) star formation and Wolf-Rayet (WR) stars trace the
recent star formation ($\sim 3-6$~Myr).  However, because of the
(relatively) low radio luminosity of these individual stars, at the
present time radio studies are largely limited to the Milky Way.  The
SKA will enable detailed investigations of
extragalactic UCH{\sc ii} regions and WR-stars for the first time.

\begin{figure}
\includegraphics[scale=0.3]{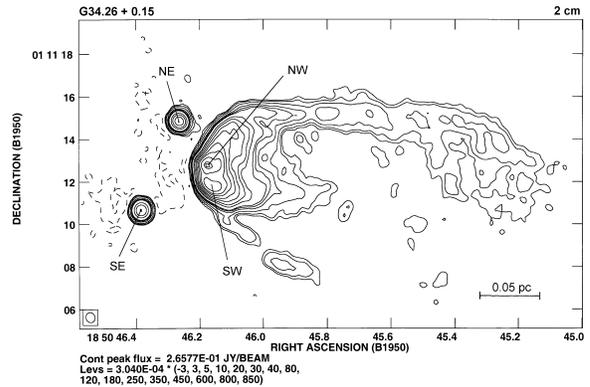}
\caption{Radio contours of the massive protostellar object
G34.26+0.15 showing a cometary ultracompact H{\sc ii} region and two
hypercompact sources to the NE and SE.  Radio recombination line
spectra show that the compact source NE has a broad line width,
while the compact source SE has a narrow line width
\cite{sewilo04}. These may correspond to different evolutionary
states. \label{G34} }
\end{figure}

Massive stars evolve through several phases that can be studied in
the radio, shedding light on the physics and time-dependence of star
formation in galaxies.  Dense prestellar cores begin their
gravitational collapse on poorly known timescales, forming hot molecular
cores with $\sim$10$^5$~year lifetimes.  Once accretion has slowed
sufficiently (possibly stopped), a hypercompact H{\sc ii} region
forms, which in turn (probably) evolves into ultracompact H{\sc ii}
region, providing the first opportunities to study the protostar as it
joins the main sequence (Figure \ref{G34}).  There is probably a very
compact H{\sc ii} region in the hot core phase, but existing radio
telescopes have had insufficient sensitivity to detect it.  Detection
by the Square Kilometer Array would be an extremely important
development in understanding the earliest, heavily accreting phase of
massive protostellar development.

Studying ultracompact H{\sc ii} regions aids our understanding of
massive star formation on the timescales of their first
$\sim$10$^5$~years.  After the ultracompact H{\sc ii} region stage of
massive stars, their main sequence lifetimes can be observed as
conventional H{\sc ii} regions for the next several Myr.  Following
the process further involves tracing the shortest-lived massive stars,
Wolf-Rayet (WR) stars with lifetimes of a few $\times 10^6$~years.
Because of their luminosities, WR stars can have a major impact on the
optical and ultraviolet spectra of galaxies, and their spectral
features are commonly used as diagnostics.  As described in
section~\ref{WRs} below, radio observations can be important for
understanding the nature and lifetimes of WR-stars and for the
interpretation of starburst galaxy spectra.

\subsection{Extragalactic Ultracompact H{\sc ii} Regions}
In the Milky Way, extremely young massive star forming regions have
been resolved into individual ultracompact H{\sc ii} (UCH{\sc ii})
regions \cite{wc89a}.  During this stage of massive star evolution,
the ionizing star is still deeply embedded in its natal cocoon, but it
has probably ceased accreting material.  Thus the UCH{\sc ii} region
phase represents the first opportunity to observe fully formed massive
stars and their impact on the surrounding interstellar medium.

As their name suggests, UCH{\sc ii} regions are small; they have
diameters of $< 0.1$~pc.  However, they do not remain small for very
long.  The H{\sc ii} region will evolve from a compact embedded state
to that of a large optically visible nebula on timescales of less than
1~Myr.  UCH{\sc ii} regions are also extremely dense with electron
densities typically well in excess of $10^4$~cm$^{-3}$, which imply
correspondingly high pressures of $P/k_B > 10^7$~cm$^{-3}$~K
\cite{churchwell02}.  With such high pressures, UCH{\sc ii} regions
must either dissipate on rapid timescales or be confined by a source
of external pressure. Indeed, it has been estimated that the lifetimes
of UCH{\sc ii} regions are roughly a hundred times longer than they
would be in the absence of significant confining pressure \cite{wc89b}.

Although the UCH{\sc ii} region phase is a key stage for studying
massive star evolution, these objects are inherently difficult to
observe due to their embedded nature, small sizes, and short
lifetimes.  Because UCH{\sc ii} regions are shrouded by many
magnitudes of visual extinction, they typically cannot be observed at
wavelengths shorter than $\sim 1-2 \mu$m.  Due to their sizes of
$<0.1$~pc, extremely high spatial resolution is required in order to
observe them.  Finally, their lifetimes of $< 1$~Myr make these
objects relatively rare, requiring large data sets in order to compile
adequate statistical samples.

The SKA will have the sensitivity to detect
individual UCH{\sc ii} regions out to the distance nearly 50~Mpc, and
the spatial resolution to unambiguously disentangle UCH{\sc ii}
regions from their surrounding environment out to the distance of
$\sim 1$~Mpc (see Figure~\ref{VLA_SKA}).  While a handful of pilot
studies have attempted to identify UCH{\sc ii} regions in the most
nearby galaxies (e.g. the Magellanic Clouds \cite{indebetouw04}), the
interpretation of these data is severely limited by the lack of
spatial resolution.  The capabilities of the SKA will enable
detailed studies of extragalactic UCH{\sc ii} regions for the first
time.

Studies of ultracompact H{\sc ii} regions are mature in our Galaxy,
but nearly unexplored other galaxies, making fundamental
discoveries by the SKA almost inevitable.
The spatial distribution, morphology, and population statistics of
embedded massive stars are fundamental properties of massive star
formation that the SKA will be able to investigate.  While
observations of UCH{\sc ii} regions in the Milky Way provide an
important case study, observations of massive star formation in other
galaxies are essential for understanding the process of massive star
formation in general.

One outstanding issue that the SKA will be
able to address is the exact location of massive star formation
relative to other protostars, density enhancements in the molecular
gas, shock fronts, and various other features of the interstellar
medium.  While massive stars may typically have proper motions of
$\sim$~a few to several km~s$^{-1}$ (e.g. the Orion Trapezium members
\cite{close03}), a significant number of ``runaway'' massive stars
exhibit high velocities of up to 200~km~s$^{-1}$ \cite{gies87}.
By the time a massive star has become optically visible, it may have
moved away from its birth location by many parsecs.  Indeed, during
the lifetime of a massive star it may travel as far as $50-100$~pc
from its original location.  

Compared to galactic scales, the distance of a few to $\sim 100$~pc is
relatively small, but it is enough to impact our understanding of
where massive stars form in a cluster (or if they were formed in a
cluster at all), and erase any signature of sequential triggering.  By
observing massive stars immediately after they are formed, it will be
possible to determine exactly what fraction of massive stars (if any)
are formed in isolation.  Moreover, observations with the SKA
will enable investigations of the nature of mass segregation in
clusters -- whether it is due to competitive accretion during the
formation stage, or if it is only due to dynamical evolution of the
stellar population.  Both of the issues of clustering and competitive
accretion are critical to understanding massive star formation.  The
clustering properties of massive stars are integral to understanding
the nature of the stellar initial mass function and also the possible
role of coalescence and encounters in massive star formation.
Likewise, determining the extent to which competitive accretion may
affect the mass of stars in a cluster has important implications for
the stellar initial mass function; competitive accretion of gas in a
proto-clusters leads to a large dynamic range in the stellar mass
function regardless of the initial masses of the molecular cores
\cite{clarke00}.

For the most nearby galaxies, roughly out to the distance of M33, it
will also be possible to assess the relationship between the size of
compact and ultracompact H{\sc ii} regions and their spectral energy
distributions.  This type of analysis will be invaluable for studying
the early stages of massive star evolution and how the ionizing star
interacts with its environment.  For example, as the UCH{\sc ii}
region evolves and begins to expand and emerge from its birth
material, its spectral energy distribution over decades in wavelength
will transform.  The exact nature of this transformation will depend
on the physical state of the surrounding interstellar medium.  For
example, as UCH{\sc ii} regions begin to evolve and expand, their
environment will transform from being optically thick to optically
thin at radio wavelengths.  As the H{\sc ii} is expanding, the
ionizing flux from the embedded massive star is also dissociating the
surrounding dust and molecular material while its winds are also
helping to clear away its natal material.  Throughout this process,
the star will become less obscured at optical wavelengths while the
radio and thermal infrared emission become fainter and eventually 
undetectable.  

\begin{figure*}[htb]
\centerline{
\includegraphics*[scale=0.85]{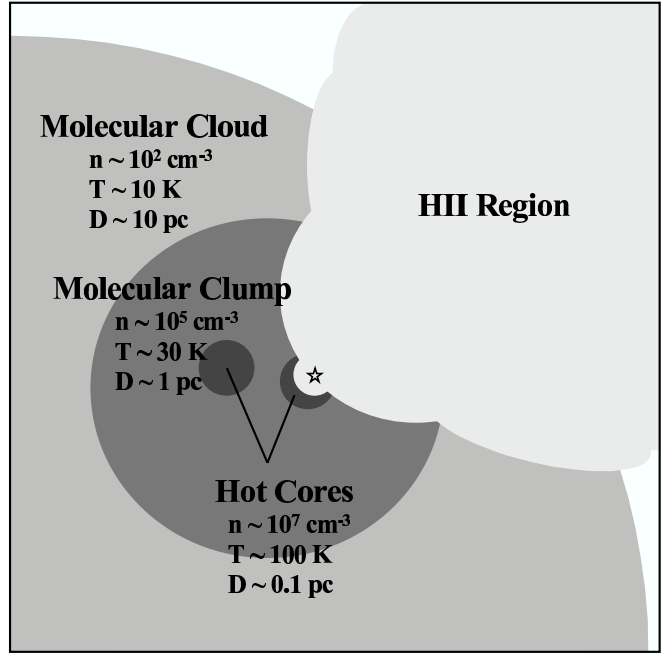}
}
\caption{A cartoon of the Kim \& Koo (2001) model for the extended halos
of emission associated with UCH{\sc ii} regions.  The Square 
Kilometer Array will be able to examine this scenario in a variety of 
nearby galaxies and environments.
 \label{Kim_model}}
\end{figure*}

In addition to simply determining the sizes of UCH{\sc ii} regions,
the SKA will also provide a tool for
investigating the leakiness and clumpiness of UCH{\sc ii} regions.
The vast majority of UCH{\sc ii} regions are not ionization bounded
(e.g. \cite{kurtz99,kim01}, and references therein), but rather are
associated with extended envelopes of emission.  This emission can
include components up to $\sim 20$~pc in size, the structure of which
will be clearly resolvable with the SKA (see
Figure \ref{Kim_model}).  Along with high spatial resolution, good
$uv$-coverage is necessary to address the ubiquity and nature of the
extended emission around UCH{\sc ii} regions.  By observing a large
sample of these objects in a variety of galaxies, the SKA will
also make it possible to ascertain whether the extended emission
depends on environmental parameters and evolutionary state.

The SKA will also enable a detailed census of the population of
ultracompact H{\sc ii} regions in a large number of galaxies.  A
comprehensive census will enable progress on a variety of issues.  For
example, the timescales which massive stars spend in the UCH{\sc ii}
region phase is rather poorly constrained by the sample of such
objects in the Milky Way.  It must be stressed that no other
diagnostics are available to ascertain the ages of UCH{\sc ii}
regions, and we are forced to rely on estimates from statistics and
dynamics.  By determining the fraction of massive stars that are still
embedded in their birth material in a large number of galaxies, it
will be possible to statistically determine the lifetime of the
UCH{\sc ii} region phase with much greater accuracy.  Moreover, by
determining the population of massive stars that are hidden from view
at optical and ultraviolet wavelengths, it will also be possible to
estimate the local and total contribution of massive stars to the
energetics of the interstellar medium in a galaxy.

The luminosity function of ultracompact H{\sc ii} regions can also
provide strong constraints on two fundamental parameters in star
formation -- the stellar and cluster initial mass functions.  The
H{\sc ii} region luminosity function is dependent on both the mass
function of the stellar cluster that powers each region and on the
distribution of mass fluctuations from which the different clusters
were formed.  Studies of the extragalactic H{\sc ii} region luminosity
function have been performed in the optical and shed light on a
variety of issues, such as whether the mass functions are different in
and out of spiral arms (e.g. \cite{keh89,mckee97}).  These studies
have not been able to probe the youngest stages of cluster formation
(compact and ultracompact H{\sc ii} regions), in which photon leakage
and evolution are least likely to be a concern.  Previous studies have
also not been able to probe the luminosity function down to the
small-numbers regime, involving single or small clusters of stars.
The SKA will allow detailed study of H{\sc ii} luminosity
functions over the full number and mass range of exciting stars, for
hundreds of diverse galaxies.

\subsection{Extragalactic Wolf-Rayet Stars \label{WRs}}

Wolf-Rayet (WR) stars are the descendants of the most massive stars
\cite{abbott87,maeder94} and exhibit powerful stellar winds with
mass-loss rates on the order of $10^{-5} - 10^{-4} M_\odot$~yr$^{-1}$
and terminal wind velocities of $\sim 10^3$~km~s$^{-1}$.  Because
massive stars have inherently short lifetimes, the WR stage of stellar
evolution happens very quickly after the onset of a starbirth event;
approximately 3-6~Myr after a burst of star formation, the massive
stars will evolve into WR stars.  Because the WR phase has such a
short duration, WR stars can provide a powerful diagnostic for tracing
how star formation progresses with time.

WR stars are visible in the radio via their strong and dense stellar
winds that produce thermal free-free emission
(e.g. \cite{seaquist76}).  One of the most fundamental parameters that
characterizes WR stars is their mass loss rate; the rate at which WR
stars lose their mass in their winds can significantly affect their
subsequent evolution and their impact on the ISM \cite{abbott80}.  The
most reliable determinations of their mass loss rates have come from
radio measurements at centimeter wavelengths because this method
requires the fewest assumptions about temperature and density
structure in the wind.  The mass loss rate can be derived from radio
observations using only three quantities \cite{wright75,panagia75} --
the distance to the star $D$, the terminal velocity of the wind
$v_\infty$, and the radio flux density $S_\nu$:
\begin{equation}
S_\nu \propto \left({{\dot{M}}\over{v_{\infty}}}\right)^{4/3} 
{{\nu^{0.6}}\over{D^2}}
\end{equation}
Once mass loss rates have been determined from radio measurements, the
velocity, temperature, and ionization structure of the wind can be
better constrained using methods available at optical and
ultraviolet wavelengths.  Moreover, because the timescales for WR star
evolution are contingent on their mass loss rates, the predicted
number of massive stars as a function of age in starburst events is
dependent on knowing these rates accurately.  This in turn affects the
interpretation of stellar synthesis models throughout the universe.

Although understanding the nature of mass loss rates from evolved
massive stars has an impact on a range of astrophysical topics
(including the interstellar medium, stellar evolution, and starburst
galaxies throughout the universe), there is not a clear
understanding of how the mass loss rates of WR stars depend on
environmental properties.  For example, because radiatively driven
winds are dependent on metal line absorption to transfer the photon
momentum, the properties of the wind must depend on metallicity.
However, the dependence on metallicity has not been well constrained
by observations; because the free-free emission scales as $\nu^{0.6}$,
stellar winds are relatively faint at radio wavelengths, and this has
prohibited radio observations of this sort in an extragalactic
context.  Furthermore, massive stars are typically formed in clusters
and embedded in H{\sc ii} regions, necessitating high spatial resolution in
order to overcome the issues of crowding and contamination.  The
SKA will enable radio observations of WR stars outside
of the Milky Way for the first time (Figure~\ref{VLA_SKA}), roughly out
to a distance of $\sim 100$~kpc.  This capability will make radio
observations of WR stars in the Magellanic Clouds possible, and
thereby vastly expand the range in environmental conditions in which 
mass loss rates from these stars can be studied.

\section{THE BIRTH OF SUPER STAR CLUSTERS}
The formation of super star clusters (SSCs) represents one of the most
extreme mode of star formation in the local universe.  With masses of
$\sim 10^4 - 10^6 M_\odot$ and radii of only a few parsecs, SSCs are
the most massive and dense stellar clusters, and it is believed that
extreme pressures are required to form them \cite{elmegreen02}.  The
properties of many SSCs are consistent with their being the
progenitors of globular clusters, although questions about their
evolution and survival remain \cite{gallagher02}.  Because of the
large number of massive stars densely packed into SSCs, these clusters
can have a violently disruptive effect on the host galaxy -- blowing
bubbles, expelling gas, enriching the interstellar and intergalactic
medium, and triggering further star formation.  The impact of massive
star clusters was probably even more important in the earlier universe
when galaxy mergers and starbursts were common, and the formation of
globular clusters was ubiquitous.  Despite the importance of massive
star clusters throughout the universe, the physical conditions
required for their formation are poorly understood.

While SSCs have been extensively studied at optical wavelengths since
the launch of the {\it Hubble Space Telescope}, very little is known
about their formation because the early stages of their evolution are
deeply enshrouded by dust.  Deep high-resolution radio observations
are required to advance this area of research.  The current radio
studies of these objects only extend out to distances of $\sim 10 -
20$~Mpc, where the sensitivity and resolution of the Very Large
Array become inadequate.  In contrast, the SKA will have the
ability to detect natal super star clusters to redshifts of nearly $z
\approx 0.1$, vastly increasing the sample of of starbursts and range
of environments in which we can study the formation of super star
clusters.  Furthermore, for the most nearby galaxies, the SKA
will resolve the physical structure of natal clusters with an
unprecedented level of detail.

\subsection{Using Radio Continuum Observations to Identify Natal Super Star 
Clusters}

\begin{figure}[t!]
\includegraphics[scale=0.3]{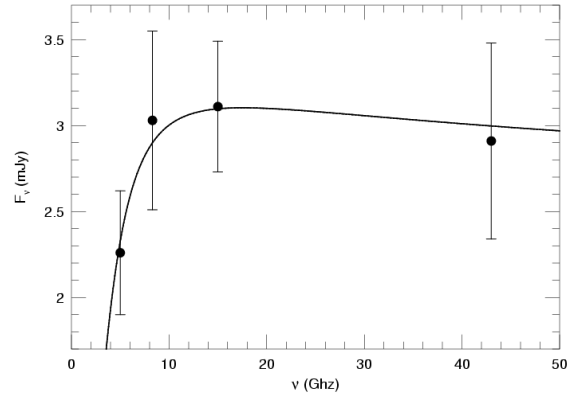}
\caption{The radio flux densities (as measured using the Very Large
Array of the most luminous natal super star cluster in the starburst
galaxy Henize~2-10 ($D\approx 10$~Mpc), along with the best fit model
H{\sc ii} region \cite{jk03}. Observations similar to these can be
used to estimate the radii and electron densities of embedded H{\sc
ii} regions, as well as the mass of the natal ionizing cluster and
pressure in the ionized gas.  \label{He2_10}}
\end{figure}

High resolution continuum observations between $\sim 5 - 20$~GHz are a
powerful way to identify natal SSCs via their ``inverted'' spectral
index ($\alpha > 0$, where $S_\nu \propto \nu^\alpha$, see
Figure~\ref{He2_10}); this kind of spectral energy distribution
results from optically thick free-free emission, similar to the
spectral energy distributions of UCH{\sc ii}s observed in the Milky
Way.  The specific spectral morphology of an H{\sc ii} region at radio
frequencies is due to a combination of size and density structure, and
these two factors can be determined given the turnover frequency and
thermal radio luminosity (as determined from observations at high
frequency where the emission is optically thin) of a cluster.  In the
case of clusters in nearby galaxies ($\lesssim 100$~Mpc), the radii of
the dense H{\sc ii} regions can be independently measured using the
high spatial resolution of the SKA.

A variety of other physical properties of the star forming regions can
also be inferred from these measurements.  The temperatures and
densities can be used to infer the pressures of the H{\sc ii} regions.
Combined with estimated cluster masses, pressures can be used to test
theories of cluster formation.  The optically-thin thermal flux
density can be used to infer a total ionizing luminosity, and the
ionizing luminosity can be translated into an embedded stellar mass by
assuming a stellar initial mass function.  These measurements can also
be used to determine the total extinction in these star forming
regions by comparing the ionized flux density as determined from radio
observations with observations of recombination lines in the near- and
mid-IR (such as Br$\gamma$, Pa$\alpha$, and Pf$\alpha$).  If the
youngest clusters are the most embedded, the extinction measure can be
used to crudely place the clusters on an evolutionary sequence.

\subsection{Using Radio Recombination Lines}

Observations of radio recombinations lines (RRLs) can be used to
constrain a range of physical properties of the ionized gas, including
electron density, temperature, gas mass, spatial structure, and
kinematics.  The major benefit of using RRLs to estimate physical
properties is that they do not suffer from the extinction that plagues
optical recombination lines (e.g. the Balmer series); therefore RRLs
can provide a powerful way to probe star forming regions.

RRLs have been detected in $\sim 20$ nearby galaxies to date
(\cite{mohan01}, and references therein) from gas with extremely high
emission measures in the nuclear regions of starburst galaxies as far
away as Arp~220 \cite{anantha00}.  However, even for the H92$\alpha$
line (8.1~GHz), which is located in the most sensitive observing band
for the Very Large Array, these kind of observations are both
difficult and expensive; typical H92$\alpha$ observations of nearby
galaxies have relatively poor velocity resolution ($\sim
100-200$~km~s$^{-1}$), spatial resolutions of only $\gtrsim 1''$
($\gtrsim 350$~pc at the distance of Arp~200), and required fairly
long integrations ($\gtrsim 6-12$ hours) to reach an rms noise in each
channel of $\gtrsim 50-100 \mu$Jy.  The sensitivity, spatial
resolution, and spectral resolution of the SKA
will allow the kinematic imaging of ionized gas in a range of
extragalactic star forming environments that are far beyond the
capabilities of the current radio facilities.

The SKA may also enable observations of radio recombination
lines from elements other than hydrogen associated with star forming
regions outside of the Milky Way.  In particular, carbon recombination
lines that are associated photo-dissociation regions (PDRs) will be
useful for probing the physical environments immediately surrounding
natal SSCs amid large and variable amounts of dust.  Because carbon is
the most abundant element with an ionization potential lower than that
of hydrogen (11.3~eV), it produces RRLs in the PDRs surrounding young
massive stars.  As with hydrogen lines, the carbon lines can be used
to constrain the physical properties of the region, including density
and temperature \cite{roshi04}.  However, the integrated intensity of
carbon RRLs are typically only a few percent of that of hydrogen
\cite{balick74}, requiring very sensitive observations.

\subsection{Using Masers}

Given the tremendous number and density of young massive stars in
natal super star clusters, it is logical to ask whether we should
expect to see other typical signposts of massive star formation
associated with these objects.  For example, molecular masers are
common in the vicinity of newly formed massive stars, and $H_2 O$
masers in particular are signposts of massive star formation
\cite{churchwell02}.  Approximately 70\% of UCH{\sc ii}s in the Milky
Way are associated with $H_2 O$ masers \cite{churchwell90}, which
suggests that the conditions required for these masers are fairly
persistent over the lifetimes of UCH{\sc ii}s.  Therefore, at a given
time, a large fraction of the massive stars in a natal SSCs should be
associated with water masers.  Because masers are typically confined
to small and dense clumps, they can probe the kinematics of star
forming regions on very small scales, providing information on both
organized and turbulent motions such as accretion and outflow.

Recent observations with large single-dish radio antennae (e.g.
Effelsberg and the Byrd Green Bank Telescope) have shown that
prominent star forming regions in nearby galaxies can be associated
with $H_2 O$ ``kilomasers''. Compared to ``megamasers'' observed in
active galaxies, the masers associated with extragalactic star
formation are relatively weak.  Water kilomasers have thus far been
detected outside of nuclear regions toward five galaxies (see
\cite{tarchi02}, and references therein).  The most distant galaxy
found to contain such extra-nuclear water masers to date is NGC~2146
(14.5~Mpc, \cite{tarchi02}).

Systematic and detailed studies of extragalactic extra-nuclear masers
associated with star formation cannot be done without a new sensitive
and high resolution radio observatory such as SKA.  The SKA will
enable us to address whether $H_2 O$ maser emission is common in
starburst galaxies at the kilomaser level and below.  With the spatial
resolution of the SKA, it will also be possible to pinpoint the
location of the maser emission, and assess the specific environments
in which this type of emission found.

\section{DETERMINING STAR FORMATION RATES}

Accurate measurement of the overall star formation rates in galaxies
is important for a broad range of astrophysical studies (see also 
van der Hulst et al. in this volume).  The star
formation rate (SFR) has direct implications for addressing the
relationship between star formation and galactic evolution, mechanical
energy input into the interstellar medium, and metal enrichment, just
to name a few issues.  A variety of techniques at different
wavelengths have been used to estimate star formation rates including
those that use the Balmer emission lines to measure the ionized gas,
the ultraviolet continuum to observe the hot young stars, observations
of infrared and sub-millimeter flux to estimate to bolometric
luminosity, and observations of the non-thermal radio continuum to
determine the synchrotron flux density due to recent supernovae
(\cite{adelberger00}, and references therein).  However, these methods
have been plagued by uncertain corrections that need to be applied in
different physical scenarios.  Ultraviolet and optical measurements
are sensitive to dust obscuration; infrared and sub-millimeter
observations are difficult to obtain due to the Earth's atmosphere and
the resulting estimates depend on the dust content and properties;
non-thermal radio techniques (typically done at 20~cm) can only
indirectly measured the SFR via cosmic ray acceleration from
supernovae.

In order to estimate star formation rates most effectively, a
technique must measure a quantity associated with young stars in a
direct and well-defined way with as few assumptions as possible.  To
date, radio studies of the SFR in galaxies have largely relied on
measuring the non-thermal flux density, which is not directly related
to star formation.  The non-thermal synchrotron flux density of a galaxy
depends critically on the magnetic field strength of a galaxy and the
cosmic ray production rate.  In particular, the synchrotron emission
from low-luminosity galaxies is often suppressed \cite{wunderlich88}.
In principle, the thermal free-free emission from a galaxy is an ideal
tracer of the SFR because it directly reflects the young massive star
content.  However, in practice it is difficult to isolate the thermal
free-free component (which has a spectral index of $\alpha = -0.1$,
where $S_\nu \propto \nu^\alpha$) from the non-thermal synchrotron
component (which typically has a steep spectral index of $\alpha \sim
-0.8$):
\begin{equation}
S_\nu = S_{syn} \nu^{\alpha} + S_{therm} \nu^{-0.1}
\end{equation}
At wavelengths longer than approximately a few cm, free-free emission is
typically fainter than the synchrotron flux density, and accurate
measurements of a galaxy's radio flux density over more a large range
in radio wavelengths are required in order to disentangle the two flux
density components.  In particular, sensitive observations at $\sim
1$~cm are necessary to disentangle the thermal flux density from the
synchrotron emission.  However, at high frequencies where the thermal
component begins to dominate the radio flux density, most normal galaxies are
inherently faint \cite{condon92}.

Once the thermal flux density is disentangled, it can be converted
into an ionizing luminosity because it directly reflects the amount of
ionized hydrogen.  Following \cite{condon92},
\begin{eqnarray}
{{Q_{Lyc}}} & \geq & 
7.9\times10^{53}~{{\rm s}^{-1}}\Big({{T_e}\over{10^4~{\rm K}}}\Big)^{-0.45}
\Big({{\nu}\over{{\rm GHz}}}\Big)^{0.1} \nonumber\\&&
\times\Big({{S_{therm} D^2}\over{10^{27}~{\rm erg~s^{-1}~Hz}^{-1}}}\Big),
\end{eqnarray}
where $D$ is the distance to the galaxy.  The inequality reflects
the possibility that a fraction of the ionizing photons could be
absorbed by dust.  This ionizing luminosity can  be converted to a
star formation rate following \cite{kennicutt98},
\begin{equation} 
{{SFR(M_\odot {\rm year}^{-1})}}
= 1.08\times 10^{-53}~Q_{Lyc}({{\rm s}^{-1}}).
\end{equation}
Using these equations as guides, the SKA will easily be able to
measure star formation rates at a redshift of $z\sim 1$ as low as
$\approx 1 M_\odot {\rm year}^{-1}$, as well as provide an independent
constraint on the star formations rates in Lyman break galaxies at $z
\sim 3$ estimated to be tens to hundreds of $ M_\odot {\rm year}^{-1}$
\cite{shapley01}.

\section{THE SQUARE KILOMETER ARRAY IN COMBINATION WITH OTHER OBSERVATORIES}
Because star formation activity is primarily observable at wavelengths
longer than the near-IR, the host of facilities that are becoming
available now and over the next decade at thermal-IR and millimeter
wavelengths are complementary to the Square Kilometer Array.
While, the Atacama Large Millimeter Array, {\it Spitzer}, {\it Stratospheric
Observatory for Infrared Astronomy}, and {\it James Webb Space
Telescope} will probe the dust cocoons and molecular properties of the
medium surrounding natal stars and clusters, only the Square
Kilometer Array will be able to observe their H{\sc ii} regions with
sufficient angular resolution.  Therefore, the combined observations
from the infrared to the radio will provide powerful and independent
diagnostics of extragalactic star forming regions.

\end{document}